\documentclass[aps,prl,floatfix,superscriptaddress,nofootinbib,
amsmath,amssymb,twocolumn,preprintnumbers]{revtex4-2}
\pdfoutput=1
\usepackage[dvipsnames]{xcolor}
\usepackage{graphicx}
\usepackage[colorlinks=true,citecolor=blue,linkcolor=blue,breaklinks=true]{hyperref}
\usepackage{soul}
\usepackage{mathtools}
\usepackage{multirow}
\usepackage{gensymb}
\usepackage{orcidlink}
\usepackage{amsmath}
\usepackage{bm}

\begin{document}

\title{
Collective neutrino-antineutrino pair oscillations}

\author{Shih-Jie Huang
\orcidlink{0009-0005-6958-7668}
}
\email{sjhuang1206@gmail.com}
\affiliation{Institute of Physics, Academia Sinica, Taipei 115, Taiwan}
\affiliation{Department of Physics, National Taiwan University, Taipei 115, Taiwan}

\author{Meng-Ru Wu
\orcidlink{0000-0003-4960-8706}}
\email{mwu@as.edu.tw}
\affiliation{Institute of Physics, Academia Sinica, Taipei 115, Taiwan}
\affiliation{Institute of Astronomy and Astrophysics, Academia Sinica, Taipei 106, Taiwan}
\affiliation{Physics Division, National Center for Theoretical Sciences, Taipei 106, Taiwan}

\begin{abstract}
In dense neutrino gas, pairing correlations between neutrinos and antineutrinos with opposite momenta can be nonzero in generalized neutrino quantum kinetic equations at the mean-field level. 
In this \emph{Letter}, we investigate for the first time the condition under which collective neutrino-antineutrino ($\nu\bar\nu$) pairing instabilities can occur, using simplified toy models consisting of discretized $\nu\bar\nu$ pairs in a homogeneous neutrino gas. 
We find that, in ansiotropic systems, $\nu\bar\nu$ pairing instabilities generally emerge when the phase space distribution of the \emph{excessive pair-occupation number}, defined as the sum of the neutrino and antineutrino occupation numbers of a pair minus 1, changes signs. 
The associated instability growth rate is set by the forward scattering potential and is comparable to that of collective fast neutrino flavor instabilities. 
The instabilities can result in pair conversions of $\nu\bar\nu$ occupation numbers between different momentum modes. 
Our results motivate further studies to assess the relevance of $\nu\bar\nu$ pairing effects in realistic astrophysical and cosmological environments.
\end{abstract}

\maketitle

\textit{\textbf{Introduction.}---}
Neutrinos play a decisive role in astrophysics and cosmology despite their weakly interacting nature~\cite{ParticleDataGroup:2024cfk,Fischer:2023ebq,DiValentino:2024xsv,Tamborra:2024fcd,Raffelt:2025wty}. 
In environments with high  neutrino densities, a variety of collective phenomena can arise, leading to nonlinear flavor evolution known as collective neutrino  oscillations; see, e.g., \cite{Duan:2010bg,Volpe:2023met,Patwardhan:2022mxg,Johns:2025mlm} for reviews. 
These effects are of significant interest due to their potential impact on astrophysical explosions such as core-collapse supernovae and binary neutron star mergers, and have been under active investigation in recent years; see,  e.g.,~\cite{Martin:2019gxb,Bhattacharyya:2020dhu,Bhattacharyya:2020jpj,Xiong:2020ntn,Capozzi:2020kge,Padilla-Gay:2020uxa,
Johns:2021qby,Li:2021vqj,Roggero:2021fyo,Nagakura:2021hyb,Wu:2021uvt,Dasgupta:2021gfs,Padilla-Gay:2021haz,Patwardhan:2021rej,Richers:2021xtf,Xiong:2021evk,
Just:2022flt,Johns:2022yqy,Xiong:2022zqz,Cervia:2022pro,Roggero:2022hpy,Shalgar:2022lvv,Illa:2022zgu,Richers:2022bkd,Xiong:2022vsy,Padilla-Gay:2022wck,Zaizen:2022cik,
Liu:2023vtz,Abbar:2023kta,Shalgar:2023ooi,Akaho:2023brj,Ehring:2023abs,Xiong:2023vcm,Martin:2023gbo,Nagakura:2023mhr,Johns:2023jjt,Johns:2023ewj,Kato:2023cig,Froustey:2023skf,Abbar:2023ltx,Nagakura:2023wbf,Nagakura:2023jfi,Balantekin:2023ayx,
Abbar:2024ynh,Johns:2024dbe,Kost:2024esc,Xiong:2024tac,Zaizen:2023ihz,Xiong:2024pue,Cornelius:2024zsb,Fiorillo:2024fnl,Fiorillo:2024qbl,Fiorillo:2024bzm,Fiorillo:2024pns,Bhaskar:2024myw,Chernyshev:2024pqy,Richers:2024zit,George:2024zxz,Lacroix:2024pbb,Siwach:2024jet,Froustey:2024sgz,Shalgar:2024gjt,Kneller:2024buy,Liu:2024nku,Goimil-Garcia:2024wgw,
Mori:2025cke,Fiorillo:2025npi,Wang:2025nii,Qiu:2025kgy,Bhattacharyya:2025gds,Lund:2025jjo,Akaho:2025giw,Dasgupta:2025quc,Fiorillo:2025zio,Froustey:2025nbi,Cornelius:2025tyt,Cornelius:2025nvd,Shi:2025ozb,Padilla-Gay:2025tko,Kost:2025vyt,Laraib:2025uza,Fiorillo:2025gkw,Xiong:2025fge,Liu:2025muc,Urquilla:2025idk,Wang:2025vbx,Goimil-Garcia:2025ozm,Qiu:2025ybw,Laraib:2025ziz,Wang:2025ihh,
Fiorillo:2026ybk,Fiorillo:2026vfo,Froustey:2026vhm,Akaho:2026kff,Carlson:2026mir,Kost:2026jrk,Urquilla:2026bff,Liu:2026roe,Fiorillo:2026byh,Zaizen:2026wvj,Fiorillo:2026tee,Viaux:2026jcy}.

In the mean-field limit, most studies have only considered one-body correlations of ultra-relativistic neutrinos and antineutrinos that are nonlinearly coupled through coherent neutrino-neutrino forward scattering~\cite{Pantaleone:1992eq,Sigl:1993ctk}.  
The spatiotemporal evolution of the neutrino and antineutrino one-body correlations is governed by the quantum kinetic equations (QKE), which includes both forward scattering potentials and momentum changing collisions~\cite{Sigl:1993ctk,Vlasenko:2013fja,blaschke2016neutrino}.

However, Refs.~\cite{Volpe:2013uxl} and \cite{Vlasenko:2013fja} pointed out separately around the same time that even at the mean-field level, nonvanishing neutrino-antineutrino ($\nu\bar\nu$) pairing correlations (between modes of opposite momenta) and helicity correlations generally exist, motivating followed-up studies that formulated the generalized QKE~\cite{Cirigliano:2014aoa,Serreau:2014cfa,Kartavtsev:2015eva}. 
Refs.~\cite{Vlasenko:2014bva,Tian:2016hec,Chatelain:2016xva,Purcell:2024bim} investigated the resonance condition for helicity correlations in astrophysical environments but found that these resonances are likely irrelevant. 

Meanwhile, the linearized stability analysis technique has been developed and widely used to study the properties of unstable modes associated with collective neutrino flavor oscillations~\cite{Banerjee:2011fj,Izaguirre:2016gsx,Capozzi:2017gqd,Yi:2019hrp,Fiorillo:2025zio}. 
Although Ref.~\cite{Vaananen:2013qja} derived the linearized equations for the generalized QKE including $\nu\bar\nu$ pairing correlations, no prior work has attempted to identify whether or not a collective pairing instability exists. 
Note that while Refs.~\cite{Sigl:1993ctk,Kartavtsev:2015eva} suggested that the pairing and helicity correlations are likely small, these are based on either heuristic arguments or steady-state assumptions. 

In this \emph{Letter}, we investigate for the first time the instability condition of $\nu\bar\nu$ pairing correlations for simplified toy models under which neutrinos and antineutrinos are represented by discrete energy and angular pairs. 
We find that $\nu\bar\nu$ pairing instabilities can occur for anisotropic systems where the phase space distribution of the \emph{excessive pair-occupation number} (EPN), defined as the sum of the neutrino and antineutrino occupation numbers of a pair minus 1, changes sign. 
We will also demonstrate that the instability leads to $\nu\bar\nu$ pair conversions from highly populated pairs to less populated ones. 
Throughout this paper, we adopt the natural units with $\hbar=c=1$. 

\textit{\textbf{Extended Quantum Kinetic Equations.}---} Assuming spatial homogeneity, neglecting the neutrino-matter forward scattering potential, taking the single-flavor scenario for simplicity, and following the notation of Ref.~\cite{Serreau:2014cfa}, the extended QKE including the $\nu\bar\nu$ pairing correlations is given by
\begin{equation}\label{eq:GQKE-compact}
i \dot{\mathcal{R}_{\bm p}}=[\mathcal{H}_{\bm p},\mathcal{R}_{\bm p}],
\end{equation}
where 
\begin{equation}\label{eq:defRH}
\mathcal{R}_{\bm p}=
\begin{pmatrix}
\rho_{\bm p} & \kappa_{\bm p} \\
\kappa_{\bm p}^\dagger & 1-\bar\rho_{\bm p} 
\end{pmatrix},~
\mathcal{H}_{\bm p}=
\begin{pmatrix}
\Gamma_{\bm p}^{\nu\nu} & \Gamma_{\bm p}^{\nu\bar\nu} \\
\Gamma_{\bm p}^{\bar\nu\nu} & \Gamma_{\bm p}^{\bar\nu\bar\nu} 
\end{pmatrix}. 
\end{equation}
In Eq.~\eqref{eq:defRH}, $\rho_{\bm p}$ and $\bar\rho_{\bm p}$ denote the occupation number of neutrinos and antineutrinos with momentum $\bm p$, and $-\bm p$, respectively, and $\kappa_{\bm p}$ is the corresponding pairing correlation. 
Inside the Hamiltonian matrix $\mathcal{H}_{\bm p}$, 
\begin{align}
\Gamma^{\nu\nu}_{\bm p}=& 
E_{\bm p} + 2g\int d^3{\bm q}
[(\rho_{\bm q}-\bar\rho_{-\bm q})(1-\hat{\bm p}\cdot\hat{\bm q}) \label{eq:Gamnn} \\
& - \hat{\bm p}\cdot (\hat\epsilon_{\bm q}\kappa_{\bm q}
+\hat\epsilon^*_{\bm q}\kappa^*_{\bm q})], \nonumber\\
\Gamma^{\bar\nu\bar\nu}_{\bm p}=& 
-E_{\bm p}+2g\int d^3{\bm q}
[(\rho_{\bm q}-\bar\rho_{-\bm q})(1+\hat{\bm p}\cdot\hat{\bm q}) \label{eq:Gambb} \\
& + \hat{\bm p}\cdot (\hat\epsilon_{\bm q}\kappa_{\bm q}
+\hat\epsilon^*_{\bm q}\kappa^*_{\bm q})], \nonumber\\
\Gamma^{\nu\bar\nu}_{\bm p}=&
-2g\hat{\epsilon}^*_{\bm p}\cdot\int d^3{\bm q}[\hat{\bm q}(\rho_{\bm q}-\bar\rho_{-\bm q})
+\hat\epsilon_{\bm q}\kappa_{\bm q}
+\hat\epsilon^*_{\bm q}\kappa^*_{\bm q}],\label{eq:Gamnb}
\end{align}
where $g\equiv\sqrt{2}G_F/(2\pi)^3$, $E_{\bm p}\simeq|\bm p|$ is the neutrino energy with momentum $\bm p$, and $\hat\epsilon_{\bm p}=\hat{\bm p}_\theta-i\hat{\bm p}_\phi$ is the polarization vector with $\hat{\bm p}_\theta$ and $\hat{\bm p}_\phi$ the corresponding unit vectors in spherical coordinates spanned by $(\hat{\bm p},\hat{\bm p}_\theta,\hat{\bm p}_\phi)$. 

Eqs.~\eqref{eq:GQKE-compact}-\eqref{eq:defRH} give rise to 
\begin{align}
& \dot \rho_{\bm p}=\dot{\bar{\rho}}_{\bm{p}}=2\rm{Im}(\Gamma_{\rm p}^{\nu\bar\nu}\kappa_{\bm p}^*),\label{eq:rhodot}\\
& i\dot \kappa_{\bm p} = (\Gamma_{\bm p}^{\nu\nu}-\Gamma_{\bm p}^{\bar\nu\bar\nu})\kappa_{\bm p}-(\rho_{\bm p}+\bar{\rho}_{\bm p}-1)\Gamma_{\bm p}^{\nu\bar\nu}.\label{eq:kappadot}
\end{align}
A few comments are in order. 
First, Eq.~\eqref{eq:rhodot} shows that the change in occupation number of $\nu$ and $\bar\nu$ with opposite momenta due to the emergence of the nonzero pairing correlation is identical. 
Second, Eq.~\eqref{eq:kappadot} indicates that for a system with an initially vanishing pairing correlation, the second term of the right hand side can generate a finite pairing correlation as with nonzero $\Gamma^{\nu\bar\nu}_{\bm p}$. 
Third, Eq.~\eqref{eq:Gamnb} indicates that anisotropy is required to generate nonzero $\Gamma_{\bm p}^{\nu\bar\nu}$.

\textit{\textbf{A Toy Model with Two Pairs.}---}We begin by examining a simplified toy model that consists of two pairs of monochromatic neutrino and antineutrino beams with the same energy $E$ traveling in perpendicular directions. 
Without loss of generality, the two neutrino beams are along the $\hat x$ and $\hat z$ directions. 
Correspondingly, the antineurino beams that can pair with the neutrino beams are set to be along the $-\hat x$ and $-\hat z$ directions. 
The (anti)neutrino occupation numbers and the pairing correlations are labeled by $\rho_x$ ($\rho_z$), $\bar\rho_x$ ($\bar\rho_z$), and $\kappa_x$ ($\kappa_z$) for the pair parallel to the $\hat x$ ($\hat z)$ direction, respectively. 
Assuming that each beam represents (anti)neutrinos occupying a phase-space volume of $\Delta V_p=(\Delta p)^3$, 
Eqs.~\eqref{eq:rhodot} and \eqref{eq:kappadot} simplify to 
\begin{align}\label{eq:rhodot_2px}
\dot\rho_x=~&\dot{\bar{\rho}}_x =
-4\mu~{\rm Im}[(\rho_z+\bar\rho_z)\kappa_x+(\kappa_z\kappa_x^*+\kappa_z\kappa_x)],\\
i\dot\kappa_x=~&
2[E-2\mu(\rho_x+\bar\rho_x)-2\mu(\kappa_z+\kappa_z^*)]\kappa_x \label{eq:kappadot_2px} \\
&-2\mu[(\rho_z+\bar\rho_z)-2\kappa_x-\kappa_z+\kappa_z^*](\rho_x+\bar\rho_x-1),\nonumber \\
\dot\rho_z=~&\dot{\bar{\rho}}_z =
4\mu~{\rm Im}[(\rho_x+\bar\rho_x)\kappa_z-(\kappa_x\kappa_z^*+\kappa_x\kappa_z)],\label{eq:rhodot_2pz}\\
i\dot\kappa_z=~&
2[E-2\mu(\rho_z+\bar\rho_z)+2\mu(\kappa_x+\kappa_x^*)]\kappa_z \label{eq:kappadot_2pz} \\
&-2\mu[-(\rho_x+\bar\rho_x)-2\kappa_z-\kappa_x+\kappa_x^*](\rho_z+\bar\rho_z-1),\nonumber
\end{align}
where $\mu=g\Delta V_p$. 
With this notation, the initial number densities of neutrinos and antineutrinos in the beams are given by $n_{\nu,x(z)}=\rho_{x(z)}^0\Delta V_p/(2\pi)^3$ and $n_{\bar\nu,x(z)}=\bar\rho_{x(z)}^0\Delta V_p/(2\pi)^3$, 
where $\rho_{x(z)}^0$ and $\bar\rho_{x(z)}^0$ are the initial occupation numbers.

Linearizing Eqs.~\eqref{eq:kappadot_2px} and \eqref{eq:kappadot_2pz} by dropping terms proportional to $\mathcal{O}(\kappa^2)$ gives rise to 
\begin{equation}\label{eq:kappa_linear}
\dot{\vec{\kappa}}=M{\vec{\kappa}}+\vec{c}. 
\end{equation}
In the above equation, 
\begin{equation}
\vec{\kappa}=
\begin{bmatrix}
\kappa_x^R \\ \kappa_x^I \\ \kappa_z^R \\ \kappa_z^I
\end{bmatrix}, 
~\vec{c}=
\mu\begin{bmatrix}
0 \\ 2\tilde\rho^0_z(\tilde\rho_x^0-1) \\
0 \\ -2\tilde\rho^0_x(\tilde\rho_z^0-1)
\end{bmatrix}, 
\end{equation}
and 
\begin{equation}
M=\begin{bmatrix}
0 & 2\tilde{E} & 0 & 4\mu(\tilde\rho^0_x-1)\\
-2\tilde{E} & 0 & 0 & 0\\
0 & 4\mu(\tilde\rho^0_z-1) & 0 & 2\tilde{E}\\
0 & 0 & -2\tilde{E} & 0
\end{bmatrix},
\end{equation}
where $\tilde{E}=E-2\mu$, $\tilde\rho^0_{x(z)}=\rho^0_{x(z)}+\bar\rho^0_{x(z)}$, and $\kappa^{R(I)}$ denotes the real (imaginary) part of $\kappa$.
The general solution of Eq.~\eqref{eq:kappa_linear} is $\vec{\kappa}(t)=e^{Mt}[\vec{\kappa}(0)+M^{-1}\vec{c}]-M^{-1}\vec{c}$.
Clearly, a positive real eigenvalue for $M$ signifies an exponential growth of pairing correlations.
The eigenvalues of $M$ are given by
\begin{equation}\label{eq:eigen_2pair}
\lambda_M=\pm 2 \sqrt{-\tilde E^2\pm2\mu\sqrt{\tilde E^2(\tilde\rho^0_x-1)(\tilde\rho^0_z-1)}}.
\end{equation}

\begin{figure}[t!]
\begin{centering}
\includegraphics[width=1.\columnwidth]{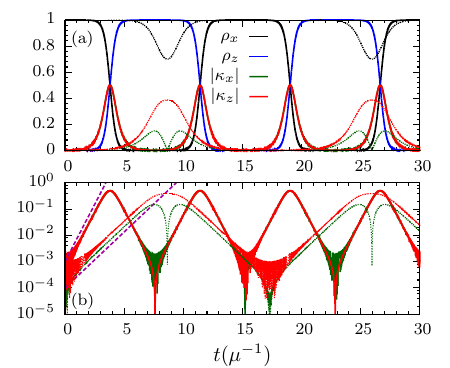}
\end{centering}
\caption{Collective $\nu\bar\nu$ pair oscillations triggered by instabilities for the toy model with two $\nu\bar\nu$ pairs traveling in the $+\hat x$ ($+\hat z$) direction. 
Panel (a): Evolution of the neutrino occupation number, $\rho_x$ ($\rho_z$), and the magnitude of the pairing correlation, $|\kappa_x|$ ($|\kappa_z|$), for $\nu\bar\nu$ pair in the $+\hat x$ ($+\hat z$) direction. 
Thick solid and thin dotted curves are for case A and case B, respectively.
Panel (B): Comparison of $|\kappa_x|$ and $|\kappa_z|$ with the prediction of their exponential growth $\propto \exp(\lambda_M^R)$ (dashed purple lines) [(see Eq.~\eqref{eq:growth_2pair})] for both case A and B.  
\label{fig:two-pair}
}
\end{figure}

With neutrinos of $\mathcal{O}(10)$MeV and typical supernova condition where $n_{\nu}\sim\mathcal{O}(10^{30-33})$~cm$^{-3}$, $E\gg\mu$.  
In this limit, $\lambda_M\sim\pm 2iE$ is pure imaginary when $(\rho^0_x+\bar\rho^0_x-1)(\rho^0_z+\bar\rho^0_z-1)>0$, representing rapidly oscillating solution with the frequency twice of the neutrino energy. 
However, when $(\rho^0_x+\bar\rho^0_x-1)(\rho^0_z+\bar\rho^0_z-1)<0$, $\lambda_M$ obtains a real part $\lambda^R_M$ given by 
\begin{equation}\label{eq:growth_2pair}
\lambda^R_M\simeq 2\mu\sqrt{|(\rho^0_x+\bar\rho^0_x-1)(\rho^0_z+\bar\rho^0_z-1)|},
\end{equation}
which leads to the unstable $\vec{\kappa}$ solution that grows exponentially. 
Note that $\lambda^R_M\sim\mathcal{O}(\mu)$ is comparable to the typical growth rate of collective fast neutrino flavor instabilities~\cite{Sawyer:2015dsa}.

We solve Eqs.~\eqref{eq:rhodot_2px}--\eqref{eq:kappadot_2pz} 
for two different sets of initial conditions. 
Case A has $\rho_x^0=\bar\rho_x^0=1$, $\rho_z^0=\bar\rho_z^0=0$, and case B has $\rho_x^0=1.0$, $\bar\rho_x^0=0.3$, $\rho_z^0=\bar\rho_z^0=0.1$.
For both cases, $\kappa_x^0=\kappa_z^0=0$, and $\mu/E=10^{-3}$.
Fig.~\ref{fig:two-pair}(a) shows the evolution of $\rho_x$, $\rho_z$, $|\kappa_x|$, and $|\kappa_z|$ for case A and $\rho_x$, $|\kappa_x|$, and $|\kappa_z|$ for case B. 
The evolutions of $\bar\rho_x$ and $\bar\rho_z$ are identical to those of $\rho_x$ and $\rho_z$ for case A, and are simply shifted by constant values for case B. 
In Fig.~\ref{fig:two-pair}(b), we show $|\kappa_x|$ and $|\kappa_z|$ for both cases together with the corresponding linearized unstable solutions $\propto\exp({\lambda_M^Rt})$. 
Clearly, even though initially $\vec{\kappa}=0$, small but nonvanishing values are being generated, followed by exponential growth with the rate $\lambda_M^R=2\mu$ (case A) and $\lambda_M^R\approx 0.490\mu$ (case B) as predicted from Eq.~\eqref{eq:growth_2pair}.
The instability leads to a very curious phenomenon that converts the initial $\nu\bar\nu$ pair along the $\hat{x}$ direction to that in the $\hat{z}$ direction. 
For case A, the maximum pairing correlation is obtained when both pairs have equal amounts of $\nu$ and $\bar\nu$. 
In this case, because $\rho_x^0=\bar\rho_x^0$, it allows full pair conversion that completely transfers the $\nu\bar\nu$ pair to the perpendicular direction. 
The evolution of system is periodic, following the ``bipolar'' or ``soliton-like'' oscillation pattern observed in certain toy models where flavor instabilities exist~\cite{Duan:2005cp,Hannestad:2006nj,Xiong:2023upa,Fiorillo:2023hlk}.
For case B, the amount of $\nu\bar\nu$ pair conversion is limited by the initial value of $\bar\rho_x=0.3$. 
Interestingly, $|\kappa_z|$ reaches maximum at the same time when most $\nu\bar\nu$ pairs along the $\hat x$ axis are converted to the $\hat z$ axis.
For other initial conditions, similar $\nu\bar\nu$ oscillations occur as long as the instability criterion is satisfied. 

Although the behavior seems peculiar, energy, momentum, and lepton numbers are all conserved. 
The only quantity that is apparently not conserved is the spin angular momentum associated with $\nu$ and $\bar\nu$. 
The initial total spin angular momentum of $\nu$ and $\bar\nu$ is along the $-\hat{x}$ direction, but is transformed into the $-\hat{z}$ direction during pair conversion. 
It is of interest to understand whether the nonzero pairing correlations carry a net angular momentum to satisfy the angular momentum conservation, or if such a process should necessarily generate a small but nonzero orbital angular momentum like in the case of particle decays~\cite{Baym:2024ptr}.

\begin{figure}[t!]
\begin{centering}
\includegraphics[width=1.\columnwidth]{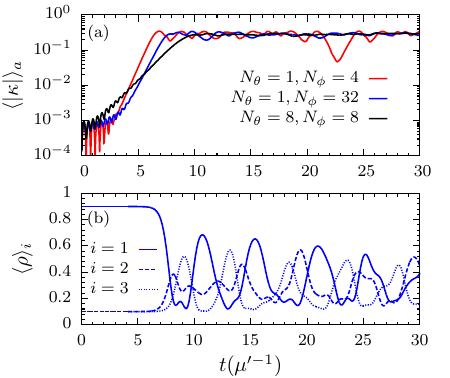}
\end{centering}
\caption{
Panel (a): Evolution of the system's averaged pairing correlation magnitude  $\langle|\kappa|\rangle_a$ for cases with multiple angular pairs. 
Panel (b): Evolution of the averaged neutrino occupation number in the $i$-th $\phi$ angular quadrant, $\langle \rho\rangle_i$. 
\label{fig:N-ang-pair}
}
\end{figure}

The analytical instability criterion $(\rho^0_x+\bar\rho^0_x-1)(\rho^0_z+\bar\rho^0_z-1)<0$ implies that this phenomenon can only occur in near-degenerate gas where the neutrino and antineutrino occupation numbers are large. 
This condition is linked to the structure of Eqs.~\eqref{eq:GQKE-compact} and \eqref{eq:defRH}. 
The diagonal components of the traceless part of $\mathcal{R}_{\bm p}$ contain only EPN ($\rho_{\bm p}+\bar\rho_{\bm p}-1$).  
Thus, EPN appears to parallel the role of the electron-minus-muon lepton occupation number (ELN) associated with collective flavor instabilities~\cite{morinaga2022fast,dasgupta2022collective,Fiorillo:2025npi}. 
This suggests that the pairing instability that we discovered here is related to the presence of the EPN crossing in anisotropic media. 
Note that for a system in thermal equilibrium (hence isotropic), EPN is always smaller than 1 for any pairs, indicating that the pairing instability can only manifest at nonequilibrium conditions.

\textit{\textbf{Beyond the Two-pair Model.}---}
While the toy model discussed above demonstrates the essence of the intriguing $\nu\bar\nu$ pair oscillation phenomenon, earlier studies investigating collective flavor oscillations for systems with imposed symmetries often found various decoherence effects when generalizing to more realistic situations with multiple angle bins~\cite{Raffelt:2007yz,Duan:2010bf}. 
In what follows, we examine $\nu\bar\nu$ pairing instabilities in systems with multiple angular or energy pairs. 

\begin{figure}[t!]
\begin{centering}
\includegraphics[width=1.\columnwidth]{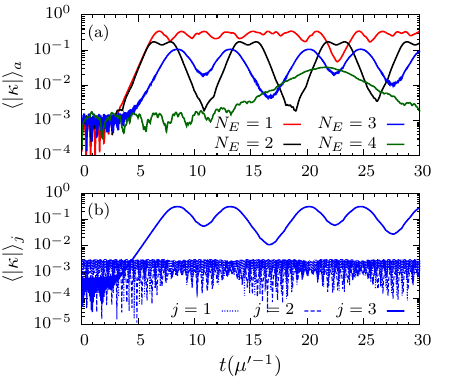}
\end{centering}
\caption{
Panel (a): Evolution of the system's averaged pairing correlation magnitude  $\langle|\kappa|\rangle_a$ for cases with multiple energy pairs. 
Panel (b): Evolution of the averaged pairing correlation magnitude in the $i$-th energy bin, $\langle |\kappa|\rangle_j$. 
\label{fig:N-eng-pair}
}
\end{figure}

For multiple pairs, we uniformly discretize the momentum space with $N_E$ energy bins for $0\leq E\leq20$~MeV, $N_\theta$ polar angle bins for $-1\leq\cos\theta\leq1$, and $N_\phi$ azimuthal angle bins for $0\leq\phi\leq 2\pi$. 
To speed up the numerical calculations, we replace $g$ in Eqs.~\eqref{eq:Gamnn}--\eqref{eq:Gamnb} by $g'=10^{-6}$~MeV$^{-2}$. 
Note that with this rescaling, the forward scattering potential is of the order $\lesssim g'\times (20{\rm MeV})^3\equiv \mu' = 0.008$~MeV, which is still $\sim \mathcal{O}(10^4)$ times smaller than the neutrino energies considered.
For $\nu\bar\nu$ pairs with $0<\phi<\pi/2$, their initial occupation numbers are $\rho^0_{E,\theta,\phi}=\bar\rho^0_{E,\theta,\phi}=0.9$. 
All other pairs have $\rho^0_{E,\theta,\phi}=\bar\rho^0_{E,\theta,\phi}=0.1$.
To speed up the calculation, we assign random perturbations between 0 and $10^{-5}$ for both real and imaginary parts of $\kappa_{E,\theta,\phi}^0$.

Fig.~\ref{fig:N-ang-pair}(a) shows the time evolution of the averaged pairing correlation of the system: $\langle |\kappa| \rangle_a = \sum_{E,\theta,\phi} |\kappa_{E,\theta,\phi}|/N_t$, where the sum runs over all pairs for  
$(N_E, N_\theta, N_\phi)=(1,1,4)$, $(1,1,32)$, and $(1,8,8)$, and $N_t=N_E N_\theta N_\phi$. 
Also shown in Fig.~\ref{fig:N-ang-pair}(b) is the evolution of the neutrino occupation number averaged over the first, second, and third quadrants in $\phi$: $\langle \rho\rangle_i = \sum_{E,\theta,\phi\in (\phi_{i,\rm min},\phi_{i,\rm max})}\rho_{E,\theta,\phi} / (N_t/4)$, where $(\phi_{i,\rm min},\phi_{i,\rm max})=(0,\pi/2)$, $(\pi/2,\pi)$, $(\pi,3\pi/2)$ for $i=1$, 2, and 3, denoting the first, second, and third $\phi$ quadrants, respectively, for $(N_E, N_\theta, N_\phi)=(1,1,32)$. 
The evolution of $\langle \rho\rangle_i$ in the fourth $\phi$ quadrant is identical to that in the second quadrant due to the imposed symmetry of the system and is not shown. 

Similarly to the two-pair toy model discussed above, the existence of the EPN crossing, together with anisotropy results in  instabilities that lead to $\nu\bar\nu$ pair conversions. 
The large $\nu$ and $\bar\nu$ occupation numbers in the first $\phi$ quadrant are being converted to other quadrants.
For all cases, $\langle |\kappa| \rangle_a$ reach values $\gtrsim\mathcal{O}(0.1)$ in the nonlinear regime.
Although not explicitly shown, we verify that the growth rate of $\langle |\kappa| \rangle_a$ in the linear (exponentially growing) regime also matches the real part of the eigenvalue of the generalized matrix $M$ for linearized and discretized Eq.~\eqref{eq:kappadot}. 
Moreover, the $\nu\bar\nu$ pair oscillations occur for all angular pairs.

For cases with multiple energy bins, the phenomenon of pair oscillations behaves differently. 
Fig.~\ref{fig:N-eng-pair}(a) shows the evolution of $\langle |\kappa| \rangle_a$ for $N_E=1,2,3$, and 4 with fixed values of $N_\theta=1$ and $N_\phi=4$.
Both the growth rate and the saturation value of $\langle |\kappa| \rangle_a$ tend to decrease as $N_E$ increases.
Interestingly, the $\nu\bar\nu$ pairing instability is largely confined within the energy bin that has the highest energy for these cases.
This is illustrated by  
Fig.~\ref{fig:N-eng-pair}(b) for $N_E=3$, where $\langle |\kappa| \rangle_j = \sum_{\theta,\phi}|\kappa_{E_j,\theta,\phi}| / (N_\theta N_\phi)$ are shown for all three $E_j$ bins.  
This is related to the fact that the normalized eigenvector of $M$ associated with the unstable eigenvalue has very small components in the subspace of all other energy bins except the one with the largest $E_j$, signifying the weak coupling between different energy modes due to $\mu' \ll E_j$, similar to the spatial Fourier mode leakage effect reported in Ref.~\cite{Abbar:2020ror}.

Although with the above setup where the initial $\nu\bar\nu$ pair distribution is uniform in the energy space, the pair instability vanishes for $N_E>4$, we have numerically found cases where the instability remains for $N_E\sim\mathcal{O}(100)$ or persists at the large $N_E$ limit by varying the initial condition. 
For instance, keeping $N_\theta=1$ and $N_\phi=4$ but assigning $\rho^0_{E,\theta,\phi}=\bar\rho^0_{E,\theta,\phi}>0.5$ \emph{only} in the $i$-th energy bins for $0<\phi<\pi/2$ while having all other pairs with $\rho^0_{E,\theta,\phi}=\bar\rho^0_{E,\theta,\phi}<0.5$, the positive solution of $\lambda_M^R\propto E_i^2\Delta E$ remains with increasing $N_E$, where $E_i$ is the energy of the $i$-th energy bin and $\Delta E$ is the bin width, i.e., the instability growth rate is proportional to the phase-space volume occupied by pairs with positive EPNs.
The same instability also remains when one increases $N_\theta$ and $N_\phi$, similarly to what discussed earlier, as long as the anisotropy remains large enough. 

While for the homogeneous setup presented here the presence of EPN crossings do not necessarily result in pairing instabilities, our preliminary analysis suggests the instabilities are shifted to inhomogeneous modes, which will be addressed in a separate wok. 
Whether or not an EPN crossing is a necessary and sufficient condition for pairing instability should be rigorously examined in the future.

\textit{\textbf{Summary and discussions.}---}
We have examined for the first time the phenomenon of $\nu\bar\nu$ pairing instability that can drive collective $\nu\bar\nu$ pair oscillations between pairs with different momenta. 
For the very simplified two-pair toy model under spatial homogeneity, we have shown that the instability associated with pairing correlations leads to oscillations of the $\nu\bar\nu$ occupation numbers of one pair to another.
We have also derived an analytical criterion showing that the instability exists when the EPNs of the two pairs take different signs. 
The corresponding growth rate in the linear regime matches perfectly with the numerical calculations.   
The growth rate is determined by forward scattering potential and is comparable to that of the collective fast neutrino flavor instabilities. 

Beyond the simplest two-pair model, we have also examined systems with multiple angular and energy bins, and found similar $\nu\bar\nu$ pairing instabilities for cases with substantial angular anisotropy and crossings in their EPN distributions. 
Interestingly, the instabilities are largely confined within one energy bin for the discretization scheme adopted here. 
Overall, these findings indicate the important role of the EPN crossing in pairing instabilities, similar to the role of ELN in flavor instabilities. 

We note that the existence of $\nu\bar\nu$ pairing instabilities discussed in this work is independent of the value of $g$.  
For the same setup but with a smaller value of $g$, the growth rate is simply proportionally smaller.  
Interestingly, we also find that at large $g$ regime (likely nonphysical for astrophysical relevance without involving physics beyond the Standard Model), instabilities that share similar properties with superconducting system in condensed matter physics can emerge. 
More dedicated studies are required to understand the underlying mechanism and possibly the associated phase boundaries. 

If the $\nu\bar\nu$ pairing instabilities discovered in this work can be realized in the densest regions in core-collapse supernovae and binary neutron star mergers, they will need to be unavoidably included in the modeling of the quantum kinetic transport of neutrinos, which can introduce extra complexity but can potentially lead to enormous consequences. 
The consideration of $\nu\bar\nu$ pairing in the many-body formulations for dense neutrino gases remains to be explored, too.
Beyond the $\nu\bar\nu$ pairing, potential instabilities associated with the helicity correlations needs to be investigated. 
We have only considered the single flavor scenario for simplicity.
Multiple flavor analysis and the potential interplay of pairing instabilities with flavor instabilities should be pursued. 
Our work presented here strongly motivates follow-up studies to elucidate not only the properties but also the practical relevance of $\nu\bar\nu$ pairing conversions in nature.

\bigskip
\textit{\textbf{Acknowledgments.---}}
We are grateful to Soumya Bhattaharyya, Jakob Ehring, Damiano Fiorillo, Hsiang-nan Li, Hiroki Nagakura, Zewei Xiong, and Di-Lun Yang for useful discussions. 
SJH and MRW acknowledge support from the National Science and Technology Council, Taiwan under Grant No.~111-2628-M-001-003-MY4, and Academia Sinica under Project No.~AS-IV-114-M04. 
MRW also acknowledges support from the Physics Division of the National Center for Theoretical Sciences, Taiwan.

\bibliographystyle{apsrev4-1}
\bibliography{main}

@article{Abbar:2024ynh,
    author = "Abbar, Sajad and Wu, Meng-Ru and Xiong, Zewei",
    title = "{Application of neural networks for the reconstruction of supernova neutrino energy spectra following fast neutrino flavor conversions}",
    eprint = "2401.17424",
    archivePrefix = "arXiv",
    primaryClass = "astro-ph.HE",
    reportNumber = "MPP-2024-12",
    doi = "10.1103/PhysRevD.109.083019",
    journal = "Phys. Rev. D",
    volume = "109",
    number = "8",
    pages = "083019",
    year = "2024"
}

@article{Shi:2025ozb,
    author = {Shi, Haihao and Huang, Zhenyang and Yan, Qiyu and Zhou, Junda and L{\"u}, Guoliang and Chen, Xuefei},
    title = "{Application of interpretable data-driven methods for the reconstruction of supernova neutrino energy spectra following fast neutrino flavor conversions}",
    eprint = "2507.09632",
    archivePrefix = "arXiv",
    primaryClass = "astro-ph.HE",
    doi = "10.1016/j.physletb.2026.140371",
    journal = "Phys. Lett. B",
    volume = "875",
    pages = "140371",
    year = "2026"
}

@article{Sawyer:2015dsa,
    author = "Sawyer, R. F.",
    title = "{Neutrino cloud instabilities just above the neutrino sphere of a supernova}",
    eprint = "1509.03323",
    archivePrefix = "arXiv",
    primaryClass = "astro-ph.HE",
    doi = "10.1103/PhysRevLett.116.081101",
    journal = "Phys. Rev. Lett.",
    volume = "116",
    number = "8",
    pages = "081101",
    year = "2016"
}

@article{Yi:2019hrp,
    author = "Yi, Changhao and Ma, Lei and Martin, Joshua D. and Duan, Huaiyu",
    title = "{Dispersion relation of the fast neutrino oscillation wave}",
    eprint = "1901.01546",
    archivePrefix = "arXiv",
    primaryClass = "hep-ph",
    doi = "10.1103/PhysRevD.99.063005",
    journal = "Phys. Rev. D",
    volume = "99",
    number = "6",
    pages = "063005",
    year = "2019"
}

@article{Capozzi:2017gqd,
    author = "Capozzi, Francesco and Dasgupta, Basudeb and Lisi, Eligio and Marrone, Antonio and Mirizzi, Alessandro",
    title = "{Fast flavor conversions of supernova neutrinos: Classifying instabilities via dispersion relations}",
    eprint = "1706.03360",
    archivePrefix = "arXiv",
    primaryClass = "hep-ph",
    reportNumber = "TIFR-TH-17-31",
    doi = "10.1103/PhysRevD.96.043016",
    journal = "Phys. Rev. D",
    volume = "96",
    number = "4",
    pages = "043016",
    year = "2017"
}

@article{Capozzi:2020kge,
    author = "Capozzi, Francesco and Chakraborty, Madhurima and Chakraborty, Sovan and Sen, Manibrata",
    title = "{Fast flavor conversions in supernovae: the rise of mu-tau neutrinos}",
    eprint = "2005.14204",
    archivePrefix = "arXiv",
    primaryClass = "hep-ph",
    reportNumber = "MPP-2020-79",
    doi = "10.1103/PhysRevLett.125.251801",
    journal = "Phys. Rev. Lett.",
    volume = "125",
    pages = "251801",
    year = "2020"
}

@article{Qiu:2025ybw,
    author = "Qiu, Yi and Radice, David and Richers, Sherwood and Guercilena, Federico Maria and Perego, Albino and Bhattacharyya, Maitraya",
    title = "{Impact of neutrino flavor conversions on neutron star merger dynamics, ejecta, nucleosynthesis, and multimessenger signals}",
    eprint = "2510.15028",
    archivePrefix = "arXiv",
    primaryClass = "astro-ph.HE",
    doi = "10.1103/qckq-78gt",
    journal = "Phys. Rev. D",
    volume = "112",
    number = "12",
    pages = "123039",
    year = "2025"
}

@article{Laraib:2025ziz,
    author = "Laraib, Zoha and Richers, Sherwood",
    title = "{Two-beam multiparticle many-body simulations of inhomogeneous fast flavor instabilities}",
    eprint = "2511.16506",
    archivePrefix = "arXiv",
    primaryClass = "astro-ph.HE",
    doi = "10.1103/jdzt-jhy2",
    journal = "Phys. Rev. D",
    volume = "112",
    number = "12",
    pages = "123045",
    year = "2025"
}

@article{Wang:2025ihh,
    author = "Wang, Tianshu and Burrows, Adam",
    title = "{The Effect of the Fast-flavor Instability on Core-collapse Supernova Models. II. Quasi-equipartition and the Impact of Various Angular Reconstruction Methods}",
    eprint = "2511.20767",
    archivePrefix = "arXiv",
    primaryClass = "astro-ph.HE",
    doi = "10.3847/1538-4357/ae3244",
    journal = "Astrophys. J.",
    volume = "997",
    number = "2",
    pages = "325",
    year = "2026"
}

@article{Liu:2026roe,
    author = "Liu, Jiabao and Nagakura, Hiroki",
    title = "{Approximate Energy-Integration Method for Identifying Collisional Neutrino Flavor Instabilities}",
    eprint = "2604.01096",
    archivePrefix = "arXiv",
    primaryClass = "astro-ph.HE",
    month = "4",
    year = "2026"
}

@article{Viaux:2026jcy,
    author = "Viaux, Nicol{\'a}s and Johns, Lucas",
    title = "{An Analytic Threshold for LESA-Driven Negative ELN Flux Directions in Core-Collapse Supernovae: Derivation and Population Census}",
    eprint = "2604.21081",
    archivePrefix = "arXiv",
    primaryClass = "astro-ph.HE",
    month = "4",
    year = "2026"
}

@article{Illa:2022zgu,
    author = "Illa, Marc and Savage, Martin J.",
    title = "{Multi-Neutrino Entanglement and Correlations in Dense Neutrino Systems}",
    eprint = "2210.08656",
    archivePrefix = "arXiv",
    primaryClass = "nucl-th",
    reportNumber = "IQuS@UW-21-034",
    doi = "10.1103/PhysRevLett.130.221003",
    journal = "Phys. Rev. Lett.",
    volume = "130",
    number = "22",
    pages = "221003",
    year = "2023"
}

@article{Chernyshev:2024pqy,
    author = "Chernyshev, Ivan and Robin, Caroline E. P. and Savage, Martin J.",
    title = "{Quantum magic and computational complexity in the neutrino sector}",
    eprint = "2411.04203",
    archivePrefix = "arXiv",
    primaryClass = "quant-ph",
    reportNumber = "IQuS@UW-21-091",
    doi = "10.1103/PhysRevResearch.7.023228",
    journal = "Phys. Rev. Res.",
    volume = "7",
    number = "2",
    pages = "023228",
    year = "2025"
}

@article{Bhaskar:2024myw,
    author = "Bhaskar, Ramya and Roggero, Alessandro and Savage, Martin J.",
    title = "{Timescales in many-body fast-neutrino-flavor conversion}",
    doi = "10.1103/PhysRevC.110.045801",
    journal = "Phys. Rev. C",
    volume = "110",
    number = "4",
    pages = "045801",
    year = "2024"
}

@article{Cervia:2022pro,
    author = "Cervia, Michael J. and Siwach, Pooja and Patwardhan, Amol V. and Balantekin, A. B. and Coppersmith, S. N. and Johnson, Calvin W.",
    title = "{Collective neutrino oscillations with tensor networks using a time-dependent variational principle}",
    eprint = "2202.01865",
    archivePrefix = "arXiv",
    primaryClass = "hep-ph",
    doi = "10.1103/PhysRevD.105.123025",
    journal = "Phys. Rev. D",
    volume = "105",
    number = "12",
    pages = "123025",
    year = "2022"
}

@article{Patwardhan:2021rej,
    author = "Patwardhan, Amol V. and Cervia, Michael J. and Balantekin, A. B.",
    title = "{Spectral splits and entanglement entropy in collective neutrino oscillations}",
    eprint = "2109.08995",
    archivePrefix = "arXiv",
    primaryClass = "hep-ph",
    reportNumber = "SLAC-PUB-17621, N3AS-21-014",
    doi = "10.1103/PhysRevD.104.123035",
    journal = "Phys. Rev. D",
    volume = "104",
    number = "12",
    pages = "123035",
    year = "2021"
}

@inbook{Patwardhan:2022mxg,
    author = "Patwardhan, Amol V. and Cervia, Michael J. and Rrapaj, Ermal and Siwach, Pooja and Balantekin, A. B.",
    editor = "Tanihata, Isao and Toki, Hiroshi and Kajino, Toshitaka",
    title = "{Many-Body Collective Neutrino Oscillations: Recent Developments}",
    booktitle = "{Handbook of Nuclear Physics}",
    eprint = "2301.00342",
    archivePrefix = "arXiv",
    primaryClass = "hep-ph",
    doi = "10.1007/978-981-15-8818-1_126-1",
    pages = "1--16",
    year = "2023"
}

@article{Balantekin:2023ayx,
    author = "Balantekin, A. Baha and Cervia, Michael J. and Patwardhan, Amol V. and Surman, Rebecca and Wang, Xilu",
    title = "{Collective Neutrino Oscillations and Heavy-element Nucleosynthesis in Supernovae: Exploring Potential Effects of Many-body Neutrino Correlations}",
    eprint = "2311.02562",
    archivePrefix = "arXiv",
    primaryClass = "astro-ph.HE",
    reportNumber = "N3AS-24-004",
    doi = "10.3847/1538-4357/ad393d",
    journal = "Astrophys. J.",
    volume = "967",
    number = "2",
    pages = "146",
    year = "2024"
}

@article{Lacroix:2024pbb,
    author = "Lacroix, Denis and Bauge, Angel and Yilmaz, Bulent and Mangin-Brinet, Mariane and Roggero, Alessandro and Balantekin, A. Baha",
    title = "{Phase-space methods for neutrino oscillations: Extension to multibeams}",
    eprint = "2409.20215",
    archivePrefix = "arXiv",
    primaryClass = "hep-ph",
    doi = "10.1103/PhysRevD.110.103027",
    journal = "Phys. Rev. D",
    volume = "110",
    number = "10",
    pages = "103027",
    year = "2024"
}

@article{Siwach:2024jet,
    author = "Siwach, Pooja and Balantekin, A. Baha and Patwardhan, Amol V. and Suliga, Anna M.",
    title = "{Exploring entanglement and spectral split correlations in three-flavor collective neutrino oscillations}",
    eprint = "2411.05169",
    archivePrefix = "arXiv",
    primaryClass = "hep-th",
    reportNumber = "FERMILAB-PUB-24-0810-T",
    doi = "10.1103/PhysRevD.111.063038",
    journal = "Phys. Rev. D",
    volume = "111",
    number = "6",
    pages = "063038",
    year = "2025"
}

@article{Xiong:2021evk,
    author = "Xiong, Zewei",
    title = "{Many-body effects of collective neutrino oscillations}",
    eprint = "2111.00437",
    archivePrefix = "arXiv",
    primaryClass = "astro-ph.HE",
    doi = "10.1103/PhysRevD.105.103002",
    journal = "Phys. Rev. D",
    volume = "105",
    number = "10",
    pages = "103002",
    year = "2022"
}

@article{Roggero:2022hpy,
    author = "Roggero, Alessandro and Rrapaj, Ermal and Xiong, Zewei",
    title = "{Entanglement and correlations in fast collective neutrino flavor oscillations}",
    eprint = "2203.02783",
    archivePrefix = "arXiv",
    primaryClass = "astro-ph.HE",
    reportNumber = "N3AS-22-006,IQuS@UW-21-023",
    doi = "10.1103/PhysRevD.106.043022",
    journal = "Phys. Rev. D",
    volume = "106",
    number = "4",
    pages = "043022",
    year = "2022"
}

@article{Roggero:2021fyo,
    author = "Roggero, Alessandro",
    title = "{Dynamical phase transitions in models of collective neutrino oscillations}",
    eprint = "2103.11497",
    archivePrefix = "arXiv",
    primaryClass = "hep-ph",
    reportNumber = "IQuS@UW-21-004",
    doi = "10.1103/PhysRevD.104.123023",
    journal = "Phys. Rev. D",
    volume = "104",
    number = "12",
    pages = "123023",
    year = "2021"
}

@article{Martin:2023gbo,
    author = "Martin, Joshua D. and Neill, Duff and Roggero, A. and Duan, Huaiyu and Carlson, J.",
    title = "{Equilibration of quantum many-body fast neutrino flavor oscillations}",
    eprint = "2307.16793",
    archivePrefix = "arXiv",
    primaryClass = "hep-ph",
    reportNumber = "LA-UR-23-28635",
    doi = "10.1103/PhysRevD.108.123010",
    journal = "Phys. Rev. D",
    volume = "108",
    number = "12",
    pages = "123010",
    year = "2023"
}

@article{Carlson:2026mir,
    author = "Carlson, Joseph and Roggero, Alessandro and Neill, Duff",
    title = "{Neutrino Flavor Evolution in High Flux Astrophysical Environments}",
    eprint = "2603.12192",
    archivePrefix = "arXiv",
    primaryClass = "hep-ph",
    reportNumber = "LA-UR-25-31666",
    month = "3",
    year = "2026"
}

@article{Lund:2025jjo,
    author = "Lund, Kelsey A. and Mukhopadhyay, Payel and Miller, Jonah M. and McLaughlin, G. C.",
    title = "{Angle-dependent in Situ Fast Flavor Transformations in Post-neutron-star-merger Disks}",
    eprint = "2503.23727",
    archivePrefix = "arXiv",
    primaryClass = "astro-ph.HE",
    reportNumber = "LA-UR-25-23002, INT-PUB-25-005, N3AS-25-005",
    doi = "10.3847/2041-8213/add0a7",
    journal = "Astrophys. J. Lett.",
    volume = "985",
    number = "1",
    pages = "L9",
    year = "2025"
}

@article{Froustey:2023skf,
    author = "Froustey, Julien and Richers, Sherwood and Grohs, Evan and Flynn, Samuel D. and Foucart, Francois and Kneller, James P. and McLaughlin, Gail C.",
    title = "{Neutrino fast flavor oscillations with moments: Linear stability analysis and application to neutron star mergers}",
    eprint = "2311.11968",
    archivePrefix = "arXiv",
    primaryClass = "astro-ph.HE",
    reportNumber = "N3AS-23-032",
    doi = "10.1103/PhysRevD.109.043046",
    journal = "Phys. Rev. D",
    volume = "109",
    number = "4",
    pages = "043046",
    year = "2024"
}

@article{Froustey:2024sgz,
    author = "Froustey, Julien and Kneller, James P. and McLaughlin, Gail C.",
    title = "{Quantum maximum entropy closure for small flavor coherence}",
    eprint = "2409.05807",
    archivePrefix = "arXiv",
    primaryClass = "hep-ph",
    reportNumber = "N3AS-24-031",
    doi = "10.1103/PhysRevD.111.063022",
    journal = "Phys. Rev. D",
    volume = "111",
    number = "6",
    pages = "063022",
    year = "2025"
}

@article{Johns:2023ewj,
    author = "Johns, Lucas",
    title = "{Neutrino many-body correlations}",
    eprint = "2305.04916",
    archivePrefix = "arXiv",
    primaryClass = "hep-ph",
    doi = "10.1142/S0217751X24501227",
    journal = "Int. J. Mod. Phys. A",
    volume = "39",
    number = "30",
    pages = "2450122",
    year = "2024"
}

@article{Kato:2023cig,
    author = "Kato, Chinami and Nagakura, Hiroki and Johns, Lucas",
    title = "{Collisional flavor swap with neutrino self-interactions}",
    eprint = "2309.02619",
    archivePrefix = "arXiv",
    primaryClass = "astro-ph.HE",
    doi = "10.1103/PhysRevD.109.103009",
    journal = "Phys. Rev. D",
    volume = "109",
    number = "10",
    pages = "103009",
    year = "2024"
}

@article{Padilla-Gay:2020uxa,
    author = "Padilla-Gay, Ian and Shalgar, Shashank and Tamborra, Irene",
    title = "{Multi-Dimensional Solution of Fast Neutrino Conversions in Binary Neutron Star Merger Remnants}",
    eprint = "2009.01843",
    archivePrefix = "arXiv",
    primaryClass = "astro-ph.HE",
    doi = "10.1088/1475-7516/2021/01/017",
    journal = "JCAP",
    volume = "01",
    pages = "017",
    year = "2021"
}

@article{Shalgar:2022lvv,
    author = "Shalgar, Shashank and Tamborra, Irene",
    title = "{Neutrino flavor conversion, advection, and collisions: Toward the full solution}",
    eprint = "2207.04058",
    archivePrefix = "arXiv",
    primaryClass = "astro-ph.HE",
    doi = "10.1103/PhysRevD.107.063025",
    journal = "Phys. Rev. D",
    volume = "107",
    number = "6",
    pages = "063025",
    year = "2023"
}

@article{Shalgar:2023ooi,
    author = "Shalgar, Shashank and Tamborra, Irene",
    title = "{Do we have enough evidence to invalidate the mean-field approximation adopted to model collective neutrino oscillations?}",
    eprint = "2304.13050",
    archivePrefix = "arXiv",
    primaryClass = "astro-ph.HE",
    doi = "10.1103/PhysRevD.107.123004",
    journal = "Phys. Rev. D",
    volume = "107",
    number = "12",
    pages = "123004",
    year = "2023"
}

@article{Fiorillo:2024pns,
    author = "Fiorillo, Damiano F. G. and Raffelt, Georg G.",
    title = "{Theory of neutrino slow flavor evolution. Part I. Homogeneous medium}",
    eprint = "2412.02747",
    archivePrefix = "arXiv",
    primaryClass = "hep-ph",
    doi = "10.1007/JHEP04(2025)146",
    journal = "JHEP",
    volume = "04",
    pages = "146",
    year = "2025"
}

@article{Fiorillo:2024bzm,
    author = "Fiorillo, Damiano F. G. and Raffelt, Georg G.",
    title = "{Theory of neutrino fast flavor evolution. Part I. Linear response theory and stability conditions.}",
    eprint = "2406.06708",
    archivePrefix = "arXiv",
    primaryClass = "hep-ph",
    doi = "10.1007/JHEP08(2024)225",
    journal = "JHEP",
    volume = "08",
    pages = "225",
    year = "2024"
}

@article{Fiorillo:2026vfo,
    author = "Fiorillo, Damiano F. G. and Raffelt, Georg G.",
    title = "{Single-wave solutions of the neutrino fast flavor system. Part II. Weak instabilities and their resonant behavior}",
    eprint = "2601.18880",
    archivePrefix = "arXiv",
    primaryClass = "hep-ph",
    month = "1",
    year = "2026"
}

@article{Fiorillo:2026ybk,
    author = "Fiorillo, Damiano F. G. and Raffelt, Georg G.",
    title = "{Single-wave solutions of the neutrino fast flavor system. Part I. Mechanical properties}",
    eprint = "2601.15372",
    archivePrefix = "arXiv",
    primaryClass = "hep-ph",
    month = "1",
    year = "2026"
}

@article{Fischer:2023ebq,
    author = "Fischer, Tobias and Guo, Gang and Langanke, Karlheinz and Martinez-Pinedo, Gabriel and Qian, Yong-Zhong and Wu, Meng-Ru",
    title = "{Neutrinos and nucleosynthesis of elements}",
    eprint = "2308.03962",
    archivePrefix = "arXiv",
    primaryClass = "astro-ph.HE",
    doi = "10.1016/j.ppnp.2024.104107",
    journal = "Prog. Part. Nucl. Phys.",
    volume = "137",
    pages = "104107",
    year = "2024"
}

@article{Tamborra:2024fcd,
    author = "Tamborra, Irene",
    title = "{Neutrinos from explosive transients at the dawn of multi-messenger astronomy}",
    eprint = "2412.09699",
    archivePrefix = "arXiv",
    primaryClass = "astro-ph.HE",
    doi = "10.1038/s42254-025-00828-2",
    journal = "Nature Rev. Phys.",
    volume = "7",
    number = "6",
    pages = "285--298",
    year = "2025"
}

@article{DiValentino:2024xsv,
    author = "Di Valentino, Eleonora and Gariazzo, Stefano and Mena, Olga",
    title = "{Neutrinos in Cosmology}",
    eprint = "2404.19322",
    archivePrefix = "arXiv",
    primaryClass = "astro-ph.CO",
    month = "4",
    year = "2024"
}

@article{ParticleDataGroup:2024cfk,
    author = "Navas, S. and others",
    collaboration = "Particle Data Group",
    title = "{Review of particle physics}",
    doi = "10.1103/PhysRevD.110.030001",
    journal = "Phys. Rev. D",
    volume = "110",
    number = "3",
    pages = "030001",
    year = "2024"
}

@article{Zaizen:2022cik,
    author = "Zaizen, Masamichi and Nagakura, Hiroki",
    title = "{Simple method for determining asymptotic states of fast neutrino-flavor conversion}",
    eprint = "2211.09343",
    archivePrefix = "arXiv",
    primaryClass = "astro-ph.HE",
    doi = "10.1103/PhysRevD.107.103022",
    journal = "Phys. Rev. D",
    volume = "107",
    number = "10",
    pages = "103022",
    year = "2023"
}

@article{Zaizen:2023ihz,
    author = "Zaizen, Masamichi and Nagakura, Hiroki",
    title = "{Characterizing quasisteady states of fast neutrino-flavor conversion by stability and conservation laws}",
    eprint = "2304.05044",
    archivePrefix = "arXiv",
    primaryClass = "astro-ph.HE",
    doi = "10.1103/PhysRevD.107.123021",
    journal = "Phys. Rev. D",
    volume = "107",
    number = "12",
    pages = "123021",
    year = "2023"
}

@article{Zaizen:2026wvj,
    author = "Zaizen, Masamichi and Nagakura, Hiroki",
    title = "{Fast Neutrino-Flavor Conversion with Attenuation and Global Lepton Gradient}",
    eprint = "2604.13617",
    archivePrefix = "arXiv",
    primaryClass = "astro-ph.HE",
    month = "4",
    year = "2026"
}

@article{Liu:2023vtz,
    author = "Liu, Jiabao and Nagakura, Hiroki and Akaho, Ryuichiro and Ito, Akira and Zaizen, Masamichi and Yamada, Shoichi",
    title = "{Universality of the neutrino collisional flavor instability in core-collapse supernovae}",
    eprint = "2310.05050",
    archivePrefix = "arXiv",
    primaryClass = "astro-ph.HE",
    doi = "10.1103/PhysRevD.108.123024",
    journal = "Phys. Rev. D",
    volume = "108",
    number = "12",
    pages = "123024",
    year = "2023"
}

@article{Akaho:2023brj,
    author = "Akaho, Ryuichiro and Liu, Jiabao and Nagakura, Hiroki and Zaizen, Masamichi and Yamada, Shoichi",
    title = "{Collisional and fast neutrino flavor instabilities in two-dimensional core-collapse supernova simulation with Boltzmann neutrino transport}",
    eprint = "2311.11272",
    archivePrefix = "arXiv",
    primaryClass = "astro-ph.HE",
    doi = "10.1103/PhysRevD.109.023012",
    journal = "Phys. Rev. D",
    volume = "109",
    number = "2",
    pages = "023012",
    year = "2024"
}

@article{Liu:2024nku,
    author = "Liu, Jiabao and Nagakura, Hiroki and Zaizen, Masamichi and Johns, Lucas and Akaho, Ryuichiro and Yamada, Shoichi",
    title = "{Quasisteady evolution of fast neutrino-flavor conversions}",
    eprint = "2411.08503",
    archivePrefix = "arXiv",
    primaryClass = "astro-ph.HE",
    doi = "10.1103/PhysRevD.111.023051",
    journal = "Phys. Rev. D",
    volume = "111",
    number = "2",
    pages = "023051",
    year = "2025"
}

@article{Akaho:2025giw,
    author = "Akaho, Ryuichiro and Nagakura, Hiroki and Yamada, Shoichi",
    title = "{Comparative testing of subgrid models for fast neutrino flavor conversions in core-collapse supernova simulations}",
    eprint = "2506.07017",
    archivePrefix = "arXiv",
    primaryClass = "astro-ph.HE",
    doi = "10.1103/37x4-jh3w",
    journal = "Phys. Rev. D",
    volume = "112",
    number = "4",
    pages = "043015",
    year = "2025"
}

@article{Kost:2025vyt,
    author = "Kost, Anson and Johns, Lucas and Duan, Huaiyu",
    title = "{Once-in-a-lifetime encounter models for neutrino media. II. Quasisteady states and miscidynamic flavor evolution}",
    eprint = "2508.14331",
    archivePrefix = "arXiv",
    primaryClass = "hep-ph",
    reportNumber = "LA-UR-25-28522",
    doi = "10.1103/x1bx-p3cm",
    journal = "Phys. Rev. D",
    volume = "112",
    number = "10",
    pages = "103004",
    year = "2025"
}

@article{Martin:2019gxb,
    author = "Martin, Joshua D. and Yi, Changhao and Duan, Huaiyu",
    title = "{Dynamic fast flavor oscillation waves in dense neutrino gases}",
    eprint = "1909.05225",
    archivePrefix = "arXiv",
    primaryClass = "hep-ph",
    doi = "10.1016/j.physletb.2019.135088",
    journal = "Phys. Lett. B",
    volume = "800",
    pages = "135088",
    year = "2020"
}

@article{Kost:2026jrk,
    author = "Kost, Anson and Duan, Huaiyu",
    title = "{Landau Damping of Collective Neutrino Oscillation Waves}",
    eprint = "2603.22246",
    archivePrefix = "arXiv",
    primaryClass = "hep-ph",
    month = "3",
    year = "2026"
}

@article{Dasgupta:2021gfs,
    author = "Dasgupta, Basudeb",
    title = "{Collective Neutrino Flavor Instability Requires a Crossing}",
    eprint = "2110.00192",
    archivePrefix = "arXiv",
    primaryClass = "hep-ph",
    reportNumber = "TIFR/TH/21-15",
    doi = "10.1103/PhysRevLett.128.081102",
    journal = "Phys. Rev. Lett.",
    volume = "128",
    number = "8",
    pages = "081102",
    year = "2022"
}

@article{Bhattacharyya:2020dhu,
    author = "Bhattacharyya, Soumya and Dasgupta, Basudeb",
    title = "{Late-time behavior of fast neutrino oscillations}",
    eprint = "2005.00459",
    archivePrefix = "arXiv",
    primaryClass = "hep-ph",
    reportNumber = "TIFR/TH/20-15",
    doi = "10.1103/PhysRevD.102.063018",
    journal = "Phys. Rev. D",
    volume = "102",
    number = "6",
    pages = "063018",
    year = "2020"
}

@article{Dasgupta:2025quc,
    author = "Dasgupta, Basudeb and Mukherjee, Dwaipayan",
    title = "{Sufficient and necessary conditions for collective neutrino instability: Fast, slow, and mixed}",
    eprint = "2505.03886",
    archivePrefix = "arXiv",
    primaryClass = "hep-ph",
    reportNumber = "TIFR/TH/25-12",
    doi = "10.1103/bqkr-rcwk",
    journal = "Phys. Rev. D",
    volume = "112",
    number = "12",
    pages = "123049",
    year = "2025"
}

@article{Xiong:2020ntn,
    author = "Xiong, Zewei and Sieverding, Andre and Sen, Manibrata and Qian, Yong-Zhong",
    title = "{Potential Impact of Fast Flavor Oscillations on Neutrino-driven Winds and Their Nucleosynthesis}",
    eprint = "2006.11414",
    archivePrefix = "arXiv",
    primaryClass = "astro-ph.HE",
    doi = "10.3847/1538-4357/abac5e",
    journal = "Astrophys. J.",
    volume = "900",
    number = "2",
    pages = "144",
    year = "2020"
}

@article{Johns:2022yqy,
    author = "Johns, Lucas and Xiong, Zewei",
    title = "{Collisional instabilities of neutrinos and their interplay with fast flavor conversion in compact objects}",
    eprint = "2208.11059",
    archivePrefix = "arXiv",
    primaryClass = "hep-ph",
    doi = "10.1103/PhysRevD.106.103029",
    journal = "Phys. Rev. D",
    volume = "106",
    number = "10",
    pages = "103029",
    year = "2022"
}

@article{Xiong:2022vsy,
    author = "Xiong, Zewei and Wu, Meng-Ru and Mart{\'\i}nez-Pinedo, Gabriel and Fischer, Tobias and George, Manu and Lin, Chun-Yu and Johns, Lucas",
    title = "{Evolution of collisional neutrino flavor instabilities in spherically symmetric supernova models}",
    eprint = "2210.08254",
    archivePrefix = "arXiv",
    primaryClass = "astro-ph.HE",
    doi = "10.1103/PhysRevD.107.083016",
    journal = "Phys. Rev. D",
    volume = "107",
    number = "8",
    pages = "083016",
    year = "2023"
}

@article{Xiong:2024tac,
    author = "Xiong, Zewei and Wu, Meng-Ru and George, Manu and Lin, Chun-Yu and Largani, Noshad Khosravi and Fischer, Tobias and Mart{\'\i}nez-Pinedo, Gabriel",
    title = "{Fast neutrino flavor conversions in a supernova: Emergence, evolution, and effects}",
    eprint = "2402.19252",
    archivePrefix = "arXiv",
    primaryClass = "astro-ph.HE",
    doi = "10.1103/PhysRevD.109.123008",
    journal = "Phys. Rev. D",
    volume = "109",
    number = "12",
    pages = "123008",
    year = "2024"
}

@article{Xiong:2025fge,
    author = "Xiong, Zewei and Wu, Meng-Ru and Largani, Noshad Khosravi and Fischer, Tobias and Mart{\'\i}nez-Pinedo, Gabriel",
    title = "{Occurrence of fast neutrino flavor conversions in QCD phase-transition supernovae}",
    eprint = "2505.19592",
    archivePrefix = "arXiv",
    primaryClass = "astro-ph.HE",
    doi = "10.1103/xl2w-dfzv",
    journal = "Phys. Rev. D",
    volume = "112",
    number = "6",
    pages = "063024",
    year = "2025"
}

@article{Xiong:2023vcm,
    author = "Xiong, Zewei and Wu, Meng-Ru and Abbar, Sajad and Bhattacharyya, Soumya and George, Manu and Lin, Chun-Yu",
    title = "{Evaluating approximate asymptotic distributions for fast neutrino flavor conversions in a periodic 1D box}",
    eprint = "2307.11129",
    archivePrefix = "arXiv",
    primaryClass = "astro-ph.HE",
    doi = "10.1103/PhysRevD.108.063003",
    journal = "Phys. Rev. D",
    volume = "108",
    number = "6",
    pages = "063003",
    year = "2023"
}

@article{Abbar:2023kta,
    author = "Abbar, Sajad",
    title = "{Applications of machine learning to detecting fast neutrino flavor instabilities in core-collapse supernova and neutron star merger models}",
    eprint = "2303.05560",
    archivePrefix = "arXiv",
    primaryClass = "astro-ph.HE",
    reportNumber = "MPP-2023-39",
    doi = "10.1103/PhysRevD.107.103006",
    journal = "Phys. Rev. D",
    volume = "107",
    number = "10",
    pages = "103006",
    year = "2023"
}

@article{Padilla-Gay:2022wck,
    author = "Padilla-Gay, Ian and Tamborra, Irene and Raffelt, Georg G.",
    title = "{Neutrino fast flavor pendulum. II. Collisional damping}",
    eprint = "2209.11235",
    archivePrefix = "arXiv",
    primaryClass = "hep-ph",
    doi = "10.1103/PhysRevD.106.103031",
    journal = "Phys. Rev. D",
    volume = "106",
    number = "10",
    pages = "103031",
    year = "2022"
}

@article{Padilla-Gay:2021haz,
    author = "Padilla-Gay, Ian and Tamborra, Irene and Raffelt, Georg G.",
    title = "{Neutrino Flavor Pendulum Reloaded: The Case of Fast Pairwise Conversion}",
    eprint = "2109.14627",
    archivePrefix = "arXiv",
    primaryClass = "astro-ph.HE",
    doi = "10.1103/PhysRevLett.128.121102",
    journal = "Phys. Rev. Lett.",
    volume = "128",
    number = "12",
    pages = "121102",
    year = "2022"
}

@article{Cornelius:2024zsb,
    author = "Cornelius, Marie and Shalgar, Shashank and Tamborra, Irene",
    title = "{Neutrino quantum kinetics in two spatial dimensions}",
    eprint = "2407.04769",
    archivePrefix = "arXiv",
    primaryClass = "astro-ph.HE",
    doi = "10.1088/1475-7516/2024/11/060",
    journal = "JCAP",
    volume = "11",
    pages = "060",
    year = "2024"
}

@article{Goimil-Garcia:2025ozm,
    author = "Goimil-Garc{\'\i}a, Manuel and Tamborra, Irene",
    title = "{Steady state of fast-oscillating neutrinos in an inhomogeneous medium}",
    eprint = "2509.22805",
    archivePrefix = "arXiv",
    primaryClass = "astro-ph.HE",
    doi = "10.1103/gdg9-rzns",
    journal = "Phys. Rev. D",
    volume = "112",
    number = "10",
    pages = "103011",
    year = "2025"
}

@article{Goimil-Garcia:2024wgw,
    author = "Goimil-Garc{\'\i}a, Manuel and Shalgar, Shashank and Tamborra, Irene",
    title = "{Pauli blocking: Probing beyond-mean-field effects in neutrino flavor evolution}",
    eprint = "2412.12268",
    archivePrefix = "arXiv",
    primaryClass = "astro-ph.HE",
    doi = "10.1103/PhysRevD.111.083054",
    journal = "Phys. Rev. D",
    volume = "111",
    number = "8",
    pages = "083054",
    year = "2025"
}










\end{document}